# IDENTIFYING 'HIDDEN' COMMUNITIES OF PRACTICE WITHIN ELECTRONIC NETWORKS: SOME PRELIMINARY PREMISES

Richard Ribeiro
rribeiro@cs.york.ac.uk

Chris Kimble
kimble@cs.york.ac.uk

University of York
Department of Computer Science
York - UK
YO10 5DD
Phone: +44 (0)1904 432-748
Fax: +44 (0)1904 432-767

*Abstract*

*This paper examines the possibility of discovering 'hidden' (potential) Communities of Practice (CoPs) inside electronic networks, and then using this knowledge to nurture them into a fully developed Virtual Community of Practice (VCoP). Starting from the standpoint of the need to manage knowledge, it discusses several questions related to this subject: the characteristics of 'hidden' communities; the relation between CoPs, Virtual Communities (VCs), Distributed Communities of Practice (DCoPs) and Virtual Communities of Practice (VCoPs); the methods used to search for 'hidden' CoPs; and the possible ways of changing 'hidden' CoPs into fully developed VCoPs. The paper also presents some preliminary findings from a semi-structured interview conducted in The Higher Education Academy Psychology Network (UK). These findings are contrasted against the theory discussed and some additional proposals are suggested at the end.*

## 1. Introduction

*Communities of Practice* (CoPs) (Lave & Wenger, 1991; Wenger, 1998; Wenger, McDermott, & Snyder, 2002) is an area of constant study and analysis. Such interest is result of a concern with methods of the creation, functioning and management of knowledge in *social communities*[1]. This subject is very relevant for many different sectors, such as industry, research institutions and businesses. Within these environments, it is important to know how to manage the knowledge generated by such communities, as their interactions can create specialized knowledge that is vital for the 'host' institution. The majority of the large companies now include knowledge (also sometimes called *social capital*) in their assets (Boersma & Stegwee, 1996).

---

[1] *Social Community* in this work represents a group of persons that participate in the same community and have active involvement in social enterprises. Participation in this sense is both personal and social, involves individual and shared feelings and is reciprocal. In addition, the members can recognize each other as belonging to the same group. [Based on (Wenger, 1998, pp. 55-56)]

The improvement in performance and the related fall in prices of computers, coupled with the spread of access to Internet in 1990s, led to an improvement in Computer-Mediated Communication (CMC) systems in general. That enhancement quickly arrived at enterprises and institutions, and later at everyone's house. As result, a new framework emerged, allowing social communities to create so-called *virtual communities*. With the creation of virtual communities came the possibility of easier 'transfer'[2] of knowledge between people in different locations - even at an international (Hiltz & Turoff, 1993; Sproull & Kiesler, 1992).

It is therefore important to examine the possibility of helping the growth of these communities, as this could make a difference for the management of the knowledge, which could influence the success of an enterprise. Companies and institutions can create an environment suitable for innovations through the facilitation of contact between groups with shared interests. Thus, allowing the nurturing of social communities that can be of use for that organisation.

These communities might be the 'seed' of an innovation that can lead to the development of new technologies, which in turn might lead to improvements in the company and institution or to the creation of new products and services. Similarly, research institutions might wish to discover potential groups and/or areas of collaboration and research as sometimes innovations are held back by a lack of communication or awareness, since the existence of similar groups inside the institution is unknown.

One step in this direction would be to discover the existence of 'hidden'[3] communities that might represent the seed of a fully developed Community of Practice (CoP). However, to accomplish this, it is necessary to analyse several related issues. First, it is necessary to be certain that those 'hidden' communities can be located. Second, it is crucial that these communities can be developed to a level of CoP, or specifically in distributed environments, to a level of a Distributed Community of Practice (DCoP). Finally, if the two previous cases are possible, it is important to verify if that DCoP can be considered a *fully developed DCoP*.

---

[2] The word *transfer* has been used between quotes because we do not believe that is possible to "transfer" knowledge in a similar way as it can be done with an object. Rather than that, we believe that the transfer is only apparent. What happens, actually, is that a person learns as the expert did (acquiring the same knowledge), as long as the conditions and environment are adequate for that.

[3] In this paper, the term *hidden* is used in the sense of potential or to-be-discovered communities. It is not our intention to discover communities that deliberately conceal themselves such as criminal groups.

## 2. A model of knowledge transfer in a distributed environment

CoPs can bring to institutions and companies novel ways to manage internal knowledge. That knowledge is shared and is formed by the sum of internal and individual knowledge. Such knowledge is also known as *tacit knowledge*, which together with *explicit knowledge*, forms the two types of knowledge a person carries (Gourlay, 2002; Nonaka, 1991, 1994; Nonaka & Takeuchi, 1995). Tacit knowledge is very hard to acquire and utterly personal. However, there are ways to 'transfer' it to others, if not in the totally, at least partially. Several authors have debated this topic; Gourlay gives a good overview of the key issues (Gourlay, 2002).

Nonaka suggested using the transfer of (tacit) knowledge as way to achieve success in companies in the early 1990's. He proposed the SECI model to explain how to accomplish the transfer of a tacit knowledge from one person to another (Nonaka, 1991, 1994; Nonaka & Takeuchi, 1995; Nonaka, Umemoto, & Senoo, 1996). The model has been widely accepted and has been used in several areas. However, some authors still argue that there are problems with it. Gourlay, for example, states that the model is not supported by empirical evidence, and that some of its phases are not coherent (Gourlay, 2003). Similarly, Jorna argues that the model lacks of background in learning theories and lacks methodology (Jorna, 1998).

The SECI model is mainly based upon the concept of apprenticeship. It explains how the tacit knowledge of an expert could be transferred to an apprentice through a process in four phases. Each phase represents a unique type of movement between tacit and explicit knowledge. Nonaka called the phases *modes of Knowledge Conversion* (Nonaka, 1994, p. 18) and the model, *the Spiral of Knowledge* (Nonaka, 1991, p. 99) or *Knowledge Spiral* (Nonaka, Umemoto, & Senoo, 1996, p. 209). The four phases are: *(S)ocialisation*, where the apprentice acquires the necessary skills working with the expert(s); *(E)xternalisation*, where the person, after acquired the tacit knowledge, transfer it to a media or pass it on; *(C)ombination*, where the knowledge (now explicit) is combined with existing explicit knowledge; and *(I)nternalisation*, where the knowledge after the previous interactions becomes new, richer and more expanded, tacit knowledge.

In addition to the move between phases, the model describes a movement in spiral, where the knowledge expands its reach (*the spiral of knowledge*). After the knowledge is individually absorbed, it goes to higher level of interaction (group, organisation, inter-organisation, etc.), where more persons can acquire it, resulting in a new internal knowledge for the company or institution.

Nonaka predicted that the SECI could be used in a distributed scenario. The spiral of knowledge, in its superior levels, is a good example that shows that the knowledge could spread in non-collated environments (e.g. between organisations). In 1994, Nonaka expanded several concepts from his first work, giving a new dimension to his previous ideas, taking in consideration the interactivity of teams and groups within companies (Nonaka, 1994). He detailed how each phase of the SECI model works in groups (potentially non-collated and/or distributed), and in doing so, he clarified the higher levels of the spiral of knowledge, showing that the SECI model could still be useful in a distributed environment.

## 3. Communities of Practice and 'hidden' CoPs

In the work of 1994, Nonaka explained that the interaction between individuals played a critical role in the development of new knowledge. He called these groups *communities of interaction* (Nonaka, 1994, p. 15). He established a relation between them and Communities of Practice, via the work of Brown and Duguid (1991). However, for better understanding of his ideas, it is first necessary to examine the main concepts behind Communities of Practice.

When initially introduced by Jean Lave and Etienne Wenger (1991), Communities of Practice were used as scenario to explain a core concept: the process of *social learning*. They discussed the idea that learning is an informal social process, rather than a planned and individual one. In this new model, learning happens mainly through social contact. The figure of the novice moves from a situation of learning in peripheral participation to full participation. The learning comes about through social interaction and observation.

The idea revealed a new realm in learning: social learning (constructivism) was used, in contrast to the behaviourism, in vogue during that period. Therefore, it seemed natural that the idea of CoPs should be explored, which was what occurred later. That first publication attracted considerable interest in different areas. It became clear that CoPs deserved a more detailed analysis. Aiming to fill that gap, Wenger released another publication (Wenger, 1998), where he conducted an detailed analysis of CoPs, giving a more scientific approach for the subject. In 2002, Wenger et al. released a third book aimed a managerial audience (Wenger, McDermott, & Snyder, 2002). This publication has a tone more practical and direct than its predecessors.

## 3.1. The main concepts related to CoPs

During this time, the concepts related to CoPs changed. Kimble and Cox have analysed this issue in (Kimble, 2006) and (Cox, 2005), respectively. Cox summarised some of those concepts and their changes over time with a table (Cox, 2005, p. 537). However, despite these alterations the main concepts remained practically the same. These concepts derive from the definition of CoPs:

*"Communities of Practice are groups of people who share a concern, a set of problems, or a passion about a topic, and who deepen their knowledge and expertise in this area by interacting on an ongoing basis."* (Wenger, McDermott, & Snyder, 2002, p. 4)

This definition outlines the main characteristics that will be present in any CoP: the *domain*, the *community* and the *practice*, which are described in (Wenger, McDermott, & Snyder, 2002):

- *Domain* - responsible for creating a sense of common identity among the members. The shared domain creates a sense of responsibility and participation in the community. It defines what the community is and is what attracts newcomers and allows them to identify themselves with it. It motivates participation, learning and gives meaning for member's actions.

- *Community* - responsible for interaction and learning among the members. The community creates a strong social bond between its participants. It motivates the improvement of the shared knowledge through joint activities and discussions, creating mutual respect and trust.
- *Practice* - represents the shared knowledge of the community. It is compounded by ideas, language, tools, frameworks and all tacit and explicit aspects of the knowledge that the community has.

The definition stands within a model with three dimensions: *mutual engagement*, *joint enterprise*, and a *shared repertoire* of experiences (Wenger, 1998, pp. 72-85). The idea is based in the assumption that, as social beings, we always engage in enterprises with persons that share a passion, mutually learning and creating, as consequence, a common knowledge. Even though a more extended and detailed set of concepts can be found in (Wenger, 1998), those listed above are the minimum requirement for a community be called *Community of Practice*.

## 3.2. Communities of Interaction and Communities of Practice

With the main concepts of CoPs explained, it is possible to return to Nonaka's statement regarding Communities of Interaction and CoPs. In 1994 Nonaka stated that Communities of Interaction play a important role in the creation of new knowledge within the organisations via the interaction of individuals that share and develop knowledge (Nonaka, 1994, p. 15).

It is now possible to say that Nonaka is, in fact, talking about Communities of Practice, as the three main elements (domain, community and practice) are all present in the Communities of Interaction. He confirms this in the same publication when he makes some comments (Nonaka, 1994, pp. 23-24) on the notion of CoPs in the work of Brown and Duguid (1991). This observation is important as it demonstrates that CoPs can be used in the same model of knowledge transfer that was described by Nonaka.

## 3.3. 'Hidden' CoPs

The benefits that CoPs can bring are, in general, only considered when the community is fully working. However, in some cases CoPs are either unknown or not established yet. In such cases, we say these communities are 'hidden', or potential communities. The 'hidden' term does not imply illegal or dangerous, it only implies that the CoPs are not yet fully visible or developed.

These 'hidden' CoPs can exist in several places within organisations, and can represent an important force. They can lead to beneficial changes inside organisations such as generating new services or products. Within educational institutions, they can result in new areas for researches or courses. However, the benefits are only applicable if those communities can become fully developed CoPs.

One characteristic of this type of latent community is that they are difficult to detect, even by people who are their potential members. That is explained by the fact of these CoPs do not yet have a physical existence. The reasons a CoP is 'hidden' can be related to several specific situations, such as the political scenario inside the organisation, lack of awareness of others with similar concerns, or even conflict of interests between the community and the organisation.

The idea of 'hidden' CoPs is not new. Since the beginning of the studies related to these communities, several suggestions were made indicating that these potential communities needed to be studied. For example, in 1991, Brown and Duguid stated: "*From our viewpoint, the central questions more involve the detection and support of emergent or existing communities*" (Brown & Duguid, 1991, p. 49). Wenger also saw these communities in 1998. He divided them in two separate states: *potential* and *latent*. The first type referring to the case where the members are somehow related. The second one concerning the case where members share past histories (Wenger, 1998, p. 228). In 2002, Wenger et al. discussed certain aspects of them (Wenger, McDermott, & Snyder, 2002, p. 70). They called such communities *loose networks*. However, they examined mainly collocated communities that already had some members connected.

Other authors have also studied potential CoPs. Cappe (2008) examines cases of latent collocated CoPs within organisations. This study makes a deep analysis of the so-called 'seeds' CoPs. Again, this study is of cases that are related to collocated CoPs at the start of their existence. There is also a body of literature that discusses the intriguing situation where CoPs *become* 'hidden' or disappear. The authors (Gongla & Rizzuto, 2004), using study cases with CoPs in IBM Global Services, discuss the reasons and characteristics of CoPs that disappeared from the organisational scene. In this study is listed the main paths followed by such communities when disappearing, the reasons why the CoPs vanish and the steps that are required to avoid or reduce it.

Searching for 'hidden' CoPs is not trivial task. The first suggestion about how this could be done might be simply to ask the members of an organisation what they like or what they care about, and later to verify if these answers can lead to the development of a fully CoP. However, as previously discussed, there may be several reasons why a community is 'hidden', and this issue needs to be carefully addressed. Thus, any research in this area needs to verify which method is appropriate for the community.

The methods that can be used might be as simple as interviews, or as complex as a longitudinal study using a combination of qualitative and quantitative methods. This is clearly a decision based in a case-to-case analysis, and is highly dependent of the organisations objectives of such search.

## 4. CoPs within electronic networks

Since the beginning of his work with CoPs, Wenger always considered the possibility of existence of non-collocated communities. In 1991, he affirmed that "*nor does the term community imply necessarily co-presence, a well-defined, identifiable group, or socially visible boundaries*" (Lave & Wenger, 1991, p. 98). In addition, in 2002 he discussed this issue deeply, dedicating an entire chapter of his book for that (Wenger, McDermott, & Snyder, 2002, p. 113). Before proceeding, it is again necessary to clarify some concepts.

### 4.1. Distributed Communities of Practice (DCoPs), Virtual Communities (VC) and Virtual Communities of Practice (VCoPs)

The term *Distributed Communities of Practice (DCoPs),* can sometimes be found in articles related to CoPs and Internet. However, its precise meaning is not always clear. The word *distributed* refers to something divided or spread. Additionally, when used together with the term *community*, it has a geographical meaning. In that case, the *distributed community* is not concentrated in a unique place, rather is divided in one or more locations. Therefore, a *Distributed Community of Practice (DCoP)* refers to a CoP that is spread over a place, or it does not have a precise delimitation of its space.

Similarly, it is easy to find publications with the term *Virtual Community (VC)*. It seems that Rheingold (1993) was the first to use the term, but after that, it is possible to find many further definitions of this term, for example, by Roberts (1998) and Igbaria (1999). In Computer Science the term evolved from the idea of something that simulates the real equivalent (e.g. *virtual memory*), to the idea of something that is real, but only exist by means of computers and networks (e.g. *virtual world*).

It can be seen that some elements are common to all definitions of Virtual Community. A general definition of *Virtual Community* based in the same common aspects might be:
*Virtual Community (VC) is the type of social community that use Computer-Mediated Communication (CMC) to keep its participants in contact.*

In the same way, with the expansion of *Computer-Mediated Communication (CMC)* and the Internet, the concept of *Distributed Communities of Practice (DCoPs)* had been reshaped to a point where became almost natural to associate *Distributed* with *Virtual*. This can be seen in great part of the publications related to CoPs at the end of 1990s and beginning of 2000s. It is now commonplace to find references only to *Virtual Communities of Practice (VCoP)*; therefore, one definition that suits this approach can be:
*Virtual Community of Practice (VCoP) is a non-collocated Community of Practice that uses Computer-Mediated Communication (CMC) to keep its participants in contact.*

## 4.2. Fully developed VCoPs

With the concept of Virtual Community of Practice (VCoP) clearly defined, it is now possible to discuss what is considered as a *fully developed VCoP*. In order to be called a fully developed VCoP, a community needs to have, at least, the following characteristics. It needs to be a:

- Social Community - the community needs to be engaged in a relationship that has active involvement in social enterprises (in accordance with the previous definition of *social community*).
- Community of Practice (CoP) - it needs to follow Wenger's definition of CoP outline previously.
- Distributed Community of Practice (DCoP) - it needs to be distributed in space in accordance with the previous definition of DCoP.
- Virtual Community (VC) - it needs to use CMC to communicate, following the previous definition of Virtual Community (VC).

### 4.3. How to change 'hidden' CoPs into fully developed VCoPs

The further development of a 'hidden' CoP is dependent on the 'discovery' of such communities. In order to do this, it is necessary to analyse several issues related to that transformation.

First, it is crucial to discover the community's desires or intentions to 'evolve' and become a fully fledge community. Wenger always highlighted passion as driving force that keeps the community together and strong (Wenger, 1998; Wenger, McDermott, & Snyder, 2002). A real CoP cannot be created by force or by any artificial means. What is possible however, is to help the development of a CoP, as Wenger detailed in (Wenger, McDermott, & Snyder, 2002). Although his advice in this publication is related to collocated CoPs, the same can be applied to VCoPs.

Another step would be to establish the forms to be used in order to help the 'hidden' CoP to flourish. This step is very dependent of the community, as only through an analysis on a case-to-case basis that it is possible to determine the best plan of action to reach that goal. It is likely that some procedures could be used in some parts of the study (e.g. the initial interviews with the potential members), but the best method(s) will only be decided after a full analysis of the community's situation. Cappe (2008) and Wenger et al. (Wenger, McDermott, & Snyder, 2002) both offer some advice on this topic.

Yet another step is to determine for how long an intervention is necessary to keep the VCoP active. This is very much related to the specification of what it is expected to achieve with the community. As a product of human interaction, the communities usually do not follow restricted rules of schedules, thus it is important to determine the limit of interference in the CoP that can be tolerated, under risk of undermining the community's self-interest. Considering all the benefits an organisation can have with a fully developed VCoP, all the procedures and time involved are worthwhile.

## 5. A small scale case study on CoPs

In order to search for 'hidden' CoPs, it is first necessary to validate the parameters used to identify existent CoPs. A small case study has already been implemented using the idea of *reification* as used in the concept of *participation-reification* duality, (Wenger, 1998, p. 57). By *reification*, we mean treating an abstraction as something with real substance or concrete existence. Under this concept, parts of a CoPs activities need to be reified by its participants during their discourse.

### 5.1. The venue

The study was applied in the *Higher Education Academic Psychology Network, UK*. The institution is one of 24 discipline-based centres within the Higher Education Academy in UK. The Psychology Network supports the teaching and learning of psychology across the UK. A core team, based at the University of York, works with staff, departments, professional bodies and overseas organisations to develop supportive networks and to improve the learning experience of psychology students in Higher Education.

### 5.2. The study

Using the concept of reification, Wenger (1998, pp. 125-126) created a list of indicators that a CoP had been formed. The list included:

1. *Sustained mutual relationships – harmonious or conflictual*
2. *Shared ways of engaging in doing things together*
3. *The rapid flow of information and propagation of innovation*

*4.   Absence of introductory preambles, as if conversations and interactions were merely the continuation of an ongoing process*

*5.   Very quick setup of a problem to be discussed*

*6.   Substantial overlap in participants' descriptions of who belongs*

*7.   Knowing what others know, what they can do, and how they can contribute to an enterprise*

*8.   Mutually defining identities*

*9.   The ability to assess the appropriateness of actions and products*

*10.  Specific tools, representations, and other artefacts*

*11.  Local lore, shared stories, inside jokes, knowing laughter*

*12.  Jargon and shortcuts to communication as well as the ease of producing new ones*

*13.  Certain styles recognized as displaying membership*

*14.  A shared discourse reflecting a certain perspective on the world*

These items verify the existence of the three main components of a CoP, as detailed in the section 3.1: *mutual engagement*, a *joint enterprise* and a *shared repertoire*. To verify the existence of the characteristics listed above, qualitative research methods were used.

A semi-structured interview was carried out, to discover the existence of 12 of the 14 items. Items 6, 8 and 14 were excluded from the interview, as they were not applicable to the chosen environment. These exclusions should not affect the overall of the research as the list is not rigid, and some items are used to verify the same characteristic. The interview was applied to the staff (7 participants) and the results are as follow.

### 5.3. Results

The answers received are listed in the table below. The results are still undergoing further analysis, but an overall picture already can be seen from the answers.

The first point to notice is that all of the answers are consistent, clearly showing the existence of a CoP in the environment. The only item that showed some diversity it is the item 4 (*"Absence of introductory preambles, as if conversations and interactions* *were merely the continuation of an ongoing process"*). That might be a result of a personal interpretation of the question. However, even in this case, the answers are in their majority the same.

| Items | Participants | | | | | | |
|---|---|---|---|---|---|---|---|
| | 1 | 2 | 3 | 4 | 5 | 6 | 7 |
| 1. | Yes | Yes | Yes | Yes | Yes | Yes | Yes |
| 2. | Yes | Yes | Yes | Yes | Yes | Yes | Yes |
| 3. | Yes | Yes | Sometimes | Yes | Yes | Yes | Yes |
| 4. | No | No | No | Sometimes | Yes | No | No |
| 5. | Yes | Yes | Yes | Yes | Yes | Yes | Yes |
| 7. | Yes | Yes | Yes | Yes | Yes | Yes | Yes |
| 9. | Yes | Yes | Yes | Yes | Yes | Yes | Yes |
| 10. | Yes | Yes | Yes | Yes | Yes | Yes | Yes |
| 11. | Yes | Yes | Yes | Yes | Yes | Yes | Yes |
| 12. | Yes | Yes | Yes | Yes | Yes | Yes | Yes |
| 13. | Yes | Yes | Yes | Yes | Yes | Yes | Yes |

Table 1: Answers for the interview

The results show that, without being aware of it, the participants have the beginning of a strong collocated CoP. Moreover, the components verified are evident. The *mutual engagement*, represented by the items 1, 7 and 9 are representative in all participants. The *joint enterprise*, represented by the items 1, 2, 4 and 5 are strong represented in the community (the exception is the question 4, discussed previously). Finally, the *shared repertoire*, represented by the items 2, 3, 10, 11, 12 and 13 are again very clear. Further study is needed to address the issue of use of CMC in the collocated CoP. Moreover, with the same type of verification is necessary in VC to discover the existence of VCoP (or maybe fully developed VCoPs).

## 6. Conclusions

The subject CoP is still full of potential, mainly because the introduction of new frameworks to create communities, for instance with the expansion of electronic networks such as the Internet. This alone can create new opportunities for practical uses of such communities in our daily life, but also within organisations: these communities could represent an agent of innovation, necessary for our modern world.

More research is necessary in this area in order to guarantee the best use of all that potential. 'Hidden' CoPs can be used to help achieving this goal, but only if several remaining issues are addressed. Additional study is necessary to understand if Virtual Communities of Practice have similar behaviour to collocated ones, and if all the original concepts and models still apply for that case. 'Hidden' CoPs are still understudied, and the search for 'hidden' VCoPs will require this knowledge; case studies are necessary in Virtual Communities to find fully developed VCoPs. Finally, more analysis is required in the issue of moving a 'hidden' CoP into a 'visible' one. In order to achieve these goals it is necessary to make more interviews in different organisations with a larger number of participants, probably using different methods and techniques, having maybe as scale for the answers or maybe even conducting an ethnographic study.

## 7. Acknowledgements

The authors are grateful to the Higher Education Academy Psychology Network for allowing its members to participate in the study case.